\documentclass[twocolumn,trackchanges,tighten]{aastex61}

\hypersetup{linkcolor=red,citecolor=blue,filecolor=cyan,urlcolor=magenta}

\usepackage{graphicx}
\usepackage{amsmath,natbib}

\newcommand{\src}{V404 Cyg}
\newcommand{\chandra}{\textsl{Chandra}}

\newcommand{\integral}{\textsl{INTEGRAL}}

\newcommand{\swift}{\textsl{Swift}}

\newcommand{\msun}{M\ensuremath{_\odot}}	
\newcommand{\mdot}{\ensuremath{\dot m}}		
\newcommand{\about}{\ensuremath{\sim}}		
\newcommand{\halpha}{H\ensuremath{\alpha}}      
\newcommand{\angstrom}{\textup{\AA}}
\newcommand{\uj}{\ensuremath{U}}		
\newcommand{\bj}{\ensuremath{B}}		
\newcommand{\vj}{\ensuremath{V}}		
\newcommand{\rc}{\ensuremath{R_{C} }}	
\newcommand{\ic}{\ensuremath{I_{C} }}	

\accepted{November 2, 2017}
\submitjournal{ApJ}

\shorttitle{WCO Monitoring of \src}
\shortauthors{Maitra et al.}

\begin{document}

\title{Simultaneous multiwavelength observations of V404 Cygni during its 2015 June outburst decay strengthen the case for an extremely energetic jet-base}


\correspondingauthor{Dipankar Maitra}
\email{maitra\_dipankar@wheatoncollege.edu}

\author[0000-0003-1897-6872]{Dipankar Maitra}
\affiliation{
Department of Physics \& Astronomy, Wheaton College,
26 East Main Street, Norton, MA 02766, USA
}

\author{John F. Scarpaci}
\affiliation{
Department of Physics \& Astronomy, Wheaton College,
26 East Main Street, Norton, MA 02766, USA
}

\author[0000-0003-2538-0188]{Victoria Grinberg}
\affiliation{
ESA European Space Research and Technology Centre (ESTEC), 
Keplerlaan 1, 2201 AZ Noordwijk, The Netherlands
}

\author[0000-0003-1621-9392]{Mark T. Reynolds}
\affiliation{
Department of Astronomy, University of Michigan,
1085 South University Avenue, Ann Arbor, MI 48109, USA
}

\author[0000-0001-9564-0876]{Sera Markoff}
\affiliation{
Anton Pannekoek Institute for Astronomy, University of Amsterdam,
1098 XH Amsterdam, The Netherlands
}

\author{Thomas J. Maccarone}
\affiliation{
Department of Physics \& Astronomy, Texas Tech University,
Lubbock, TX 79409-1051, USA
}

\author{Robert I. Hynes}
\affiliation{
Department of Physics \& Astronomy, Louisiana State University,
Baton Rouge, LA 70803, USA
}

\begin{abstract}
We present results of multiband optical photometry of the black
hole X-ray binary system \src\ obtained using Wheaton College
Observatory's 0.3m telescope, along with strictly simultaneous
\integral\ and \swift\ observations during 2015 June 25.15--26.33
UT, and 2015 June 27.10--27.34 UT. These observations were made
during the 2015 June outburst of the source when it was going through
an epoch of violent activity in all wavelengths ranging from radio
to $\gamma$-rays. The multiwavelength variability timescale favors
a compact emission region, most likely originating in a jet outflow,
for both observing epochs presented in this work. The simultaneous
\integral/Imager on Board the Integral Satellite (IBIS) 20--40 keV
light curve obtained during the June 27 observing run correlates
very strongly with the optical light curve, with no detectable delay
between the optical bands as well as between the optical and hard
X-rays. The average slope of the dereddened spectral energy
distribution was roughly flat between the \ic- and \vj-bands during
the June 27 run, even though the optical and X-ray flux varied by
$>$25$\times$ during the run, ruling out an irradiation origin for
the optical and suggesting that the optically thick to optically
thin jet synchrotron break during the observations was at a frequency
larger than that of \vj-band, which is quite extreme for X-ray
binaries.  These observations suggest that the optical emission
originated very close to the base of the jet.  A strong \halpha\
emission line, probably originating in a quasi-spherical nebula
around the source, also contributes significantly in the \rc-band.
Our data, in conjunction with contemporaneous data at other wavelengths
presented by other groups, strongly suggest that the jet-base was
extremely compact and energetic during this phase of the outburst.
\end{abstract}


\section{Introduction} \label{s:intro}
Outbursts of transient X-ray binary (XRB) systems offer unique
opportunities to study accretion flows and outflows near compact
stellar remnants such as black holes and neutron stars. Given that
the duration of an XRB outburst is typically weeks to months, during
which the luminosity can vary by many orders of magnitude, dense
multiwavelength observations during such outbursts offer a window
through which to study poorly understood aspects of accretion
processes in strong-field gravity such as changes in the accretion
flow and the flow's geometry, dominant emission mechanisms, and
disk--jet coupling.

The XRB system \src, located at a parallax distance of $2.39\pm0.14$
kpc \citep{MJ+2009}, harbors a $12^{+3}_{-2}$ \msun\ black hole
accretor and a $0.7^{+0.3}_{-0.2}$ \msun\ donor of K3 III spectral
type \citep{Shahbaz+1994, Khargharia+2010}. The orbital period of
the system is 6.5 days \citep{Casares+1992} and near-infrared (NIR)
spectroscopic observations by \citet{Khargharia+2010} indicate an
orbital inclination of $\sim$$67\degr$ with respect to our line of
sight \citep[however, see also][for a discussion on how changes in
the quiescent light curve morphology and/or uncertainty in the
disk-to-total flux ratio can affect mass and inclination
measurements]{Cantrell+2010}.  The orbital period of \src\ is
significantly larger than that of most black hole XRBs discovered
so far \citep[see, e.g.][for recent compilations of properties of
XRBs]{CS+2016,BT+2016}. It has been speculated that the large orbital
period and consequently large size of the accretion disk may at
least be partly responsible for the violent outbursts of \src\
\citep{Kimura+2016}. In contrast, outbursts of typical transient
XRB systems go through a fast-rise and exponential-decay (FRED)
light curve morphology \citep[see][for a review]{CSL1997} that is
thought to be driven by instabilities in the accretion disk
\citep{Lasota2001}.

After \about26 years of inactivity \citep{Rana+2016} \src\ went
through a violent outburst starting 2015 mid-June, when enhanced
hard X-ray activity was noticed by the Burst Alert Telescope (BAT)
on board the \swift\ mission \citep{gcn17929}. Reports of detections
at other wavelengths spanning from radio to $\gamma$-rays followed
soon after, triggering pointed observations using numerous space-borne
as well as ground-based telescopes \citep[see][for a list of circulars
and telegrams covering this outburst]{atel7959}\footnote{An expanded
list of multi-wavelength observations made during the outburst is
available at \url{http://deneb.astro.warwick.ac.uk/phsaap/v404cyg/data/}}.

As in the previous outburst of \src\ in 1989, strong and rapid
variability at all wavelengths was observed during the 2015 June
outburst, manifested by dramatic changes in flux within timescales
of minutes to seconds.  Hard X-ray flares with luminosities exceeding
6 Crab were observed with {\it INTEGRAL} \citep{Rodriguez+2015}.
Analyzing the {\it INTEGRAL} observations made between 2015 June
17--20, \citet{Natalucci+2015} and \citet{Roques+2015} reported
highly variable Compton-upscattered hard X-ray emission with seed
photons that likely originated in a synchrotron-driven jet. Analysis
of {\it Fermi} Gamma-Ray Burst Monitor (GBM) data from \src\ during
the outburst by \citet{Jenke+2016}, where the source reached up to
30 Crab with emission detected as high as 300 keV, also supports a
jet component in the observed emission.  \citet{Walton+2017}, using
{\it NuSTAR} observations, established that the X-ray emission
region should be quite compact.  Their modeling suggests that an
intense source of X-rays lies only a few ($<$10) gravitational radii
above the black hole, and that this region serves as a ``lamp post''
illuminating the accretion disk below to produce the strong reflection
features.  They interpret this as likely indicating that the X-rays
come from a jet.  This is certainly a viable possibility, but we
emphasize that there is no direct evidence that the X-ray emission
comes from a bipolar outflow, rather than a compact inflowing region.
Modeling of the 0.6--250 keV X-ray spectrum obtained on MJD 57194
from strictly simultaneous {\it Swift} and {\it INTEGRAL} observations,
\citet{Motta+2017} supported strong reflection as well. \citet{Motta+2017}
also argued for significant absorption and obscuration, akin to
spectral features seen in highly accreting but obscured active
galactic nuclei.  Analyzing 602 spectra in the energy range 20--200
keV obtained by \integral\ Imager on Board the Integral Satellite
(IBIS/ISGRI) throughout the outburst, \citet{SF+2017} found evidence
for variable absorption as well, with the absorbing column becoming
Compton thick between flares.  These authors conclude that the
flaring activity may not be solely due to intrinsic variability of
the emission source, but may also be due to changes in the absorbing
column and/or scattering.  We note that such highly variable
absorption is also atypical of most XRBs, and perhaps seen only in
sources like \src\ and V4641 Sgr \citep[see, e.g.][]{MB2006} which
exhibit extremely violent activity during outbursts.

At higher energies, \citet{epline} reported detection of a positron
annihilation line near 511 keV with {\it INTEGRAL}, suggesting a
compact, energetic region where the optical depth to pair production
was non-negligible.  The {\it AGILE} satellite detected 50--400 MeV
$\gamma$-ray emission from \src\ between MJD~57197.25--57199.25 at
$\sim$4.3$\sigma$ level \citep{Piano+2017}.  An excess 0.1--100 GeV
$\gamma$-ray emission from the direction of \src\ by the {\it Fermi}
Large Area Telescope (LAT) was detected on MJD 57199 \citep{Loh+2016}.
These $\gamma$-ray detections strengthen the argument for the
existence of a jet in \src. At the other end of the electromagnetic
spectrum, the source was highly active in radio through sub-mm
wavelengths as well \citep[see e.g.][and references therein]{AT+2017}.
Multiple Jansky-level flares were observed during a 4 hr observing
window on MJD~57195, in eight frequency bands ranging from 5 GHz
to 666 GHz.  Simultaneous modeling of these radio light curves
suggests multiple ejection events at relativistic speeds \citep{AT+2017}.

Softer ($<$10 kev) X-ray spectra obtained using {\it Chandra} HETGS
on 2015 June 22 and 23 showed strong, narrow lines due to various
species of Mg, Si, S, and Fe \citep{King+2015}. Analyzing these
lines, \citet{King+2015} inferred that the central engine is obscured
and the outer accretion disk is illuminated.  As noted above,
modeling of the broadband X-ray spectra during the outburst also
supports variable, and often Compton-thick, absorption \citep{SF+2017,
Motta+2017}.  Additionally, \citet{King+2015} noted the presence
of strong disk winds, based on observed P-Cygni profiles, during
epochs of highest flux.  P-Cygni profiles were observed not only
in soft X-rays, but also in optical spectra obtained during this
outburst, suggesting a strong outflow from the cooler, outer accretion
disk \citep{M-D+2016}. Based on estimates of the mass-loss rate due
to the wind, \citet{M-D+2016} have suggested that the brief but
violent epochs of activity of XRB transients with large accretion
disks (such as \src, V4641~Sgr, GRS~1915+105) may be regulated by
these strong winds, which can deplete the disk mass and bring the
outburst to a halt. Analyzing \swift/X-Ray Telescope (XRT) + BAT
data in the 0.5--150 keV range, \citet{Radhika+2016} also inferred
the presence of a wind outflow and a possible short-term depletion
of the inner accretion disk before a massive radio flare.

Optical spectra obtained during the outburst by \citet{M-D+2016},
\citet{Gandhi+2016}, and \citet{Rahoui+2017} indicate that the
equivalent widths of the optical emission lines became extremely
large after MJD~57200. In particular, \citet{M-D+2016} reported
\halpha\ equivalent widths up to $\sim$2000 \angstrom\ on MJD 57201,
and \citet{Rahoui+2017} reported an \halpha\ equivalent width of
1129\angstrom\ on MJD 57200. Furthermore, a multitude of emission
lines from different species appeared after MJD~57200, and the
\halpha-to-H\ensuremath{\beta} flux ratio increased from $\sim$2.5
to $\sim$6. These changes in the optical spectra led
\citet{M-D+2016} to suggest that a short-lived nebular phase was
triggered around this time.

\citet{Tanaka+2016} observed \src\ at various epochs between MJD
57190--57194 and did not find any evidence of intrinsic optical or
NIR linear polarization, or any significant change in the
amount of polarization or the polarization position angle. However,
observations made between MJD~57191.62--57191.84 using the MASTER
robotic telescope network showed a 4\%--6\% change in the optical
linear polarization \citep{Lipunov+2016}. As shown in Fig.~1 of
\citet{Lipunov+2016}, the polarization change is anticorrelated
with flux change. A similar, though smaller ($\about$1\%), change
in the amount of polarization was also noted by \citet{Shahbaz+2016}
between MJD~57197.12 and 57197.2.

While the most active phase of this outburst lasted only for about
a couple of weeks, low-level activity continued for a couple of
months.  Based on \swift\ monitoring data, \citet{atel7959} estimated
that the source activity declined back to quiescence sometime between
2015 August 5 and August 21. The radio and X-ray decay of this
outburst into quiescence has been studied by \citet{Plotkin+2017}.

In the context of these results, we were interested in understanding
the origin of the optical emission.  While the optical emission
during high \mdot\ (mass accretion rate) epochs of a typical XRB
outburst is usually thermal-dominated, a non-thermal component
sometime appears during epochs of low \mdot\ and hard X-ray spectral
state \citep{Russell+2013b}.  In sources such as GRS~1915+105 and
XTE J1550-564, the optical and NIR occasionally also reveals
the presence of a non-thermal component during extreme flaring
events \citep{Buxton+2012, Russell+2010}.  In this paper we focus
on ground-based optical observations made in the \vj-, \rc-, and \ic-bands
during 2015 June 25 and 27.  We also present the \integral/IBIS
(20-40 keV), \swift/XRT (0.3-10 keV), and \swift/Ultraviolet and Optical
Telescope (UVOT) U-band light curves that are strictly simultaneous with 
our observations, and the corresponding
time-series auto- and cross-correlation analyses of our ground-based
data as well as \integral\ data. The observation details and data
reduction process are detailed in \S\ref{s:data}, data analysis is
presented in \S\ref{s:analysis}, and we discuss our conclusions in
\S\ref{s:conclusion}.

\begin{figure*}[!ht] 
\centering
 \includegraphics[height=0.99\textwidth, angle=-90, clip, trim=0mm 0mm -4mm 0mm]
 {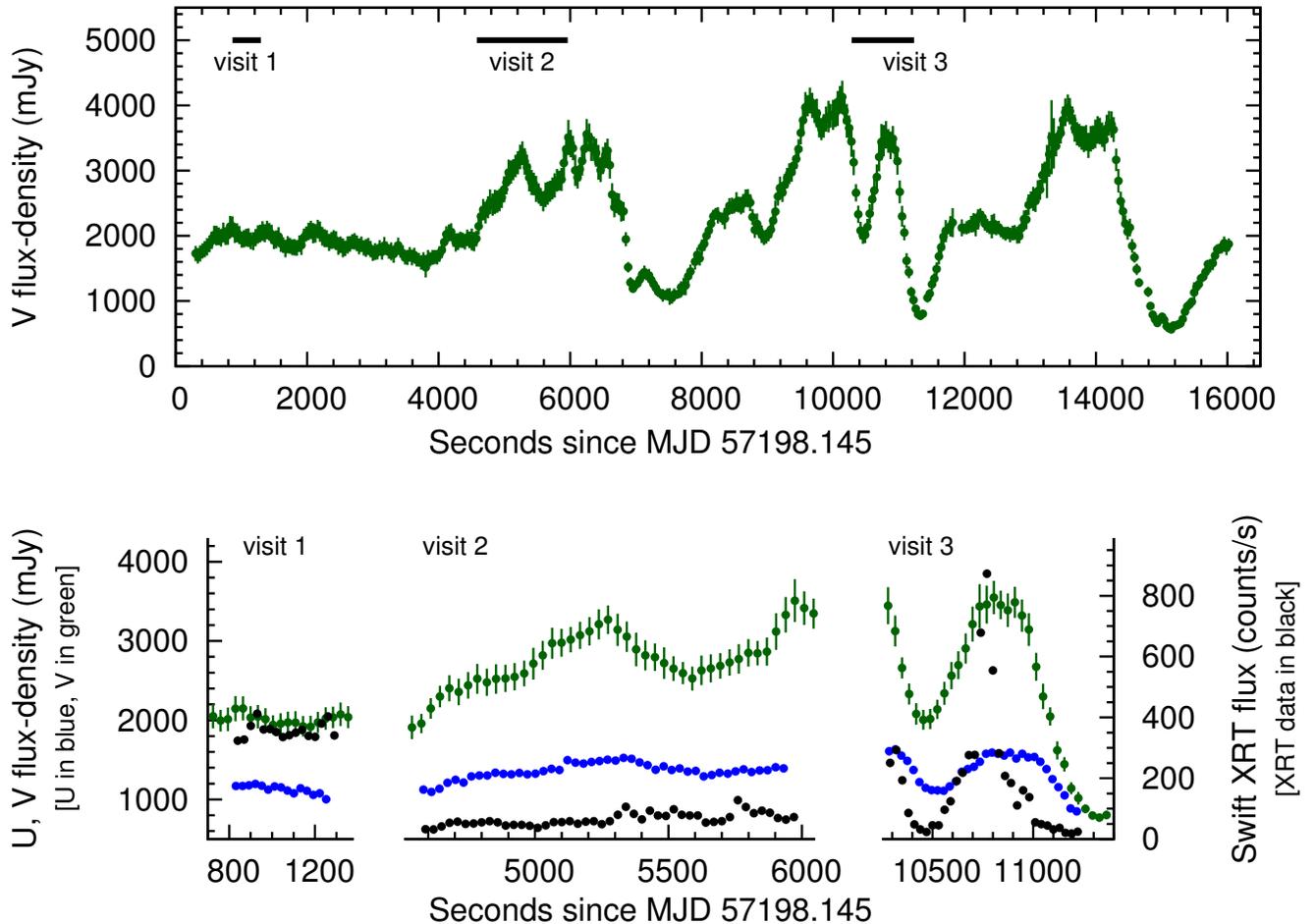} 
   \caption{
   Dereddened, interloper-subtracted \vj-band light curve obtained
   by the WCO 0.3m telescope on MJD 57198 is shown in green in both
   the top and bottom panels. While the top panel shows the \vj-band
   light curve for the entire night, the bottom three subpanels zoom
   in to show the \swift/UVOT U-band (in blue) and \swift/XRT (in
   black) light curves as well.  The ordinate ranges are the same
   in all three bottom subpanels to allow easy comparison of flux
   variability among the three \swift\ visits.
	{\it The WCO and \swift\ light curve data are included with 
	this published article as data behind the figure (DbF) in 
	machine-readable table format.}
   }
 \label{f:n1-lc} 
\end{figure*} 

\begin{figure*}[!ht] 
\centering
 \includegraphics[width=0.85\textwidth,
	angle=-90,clip,trim=0mm 0mm -5mm 0mm]{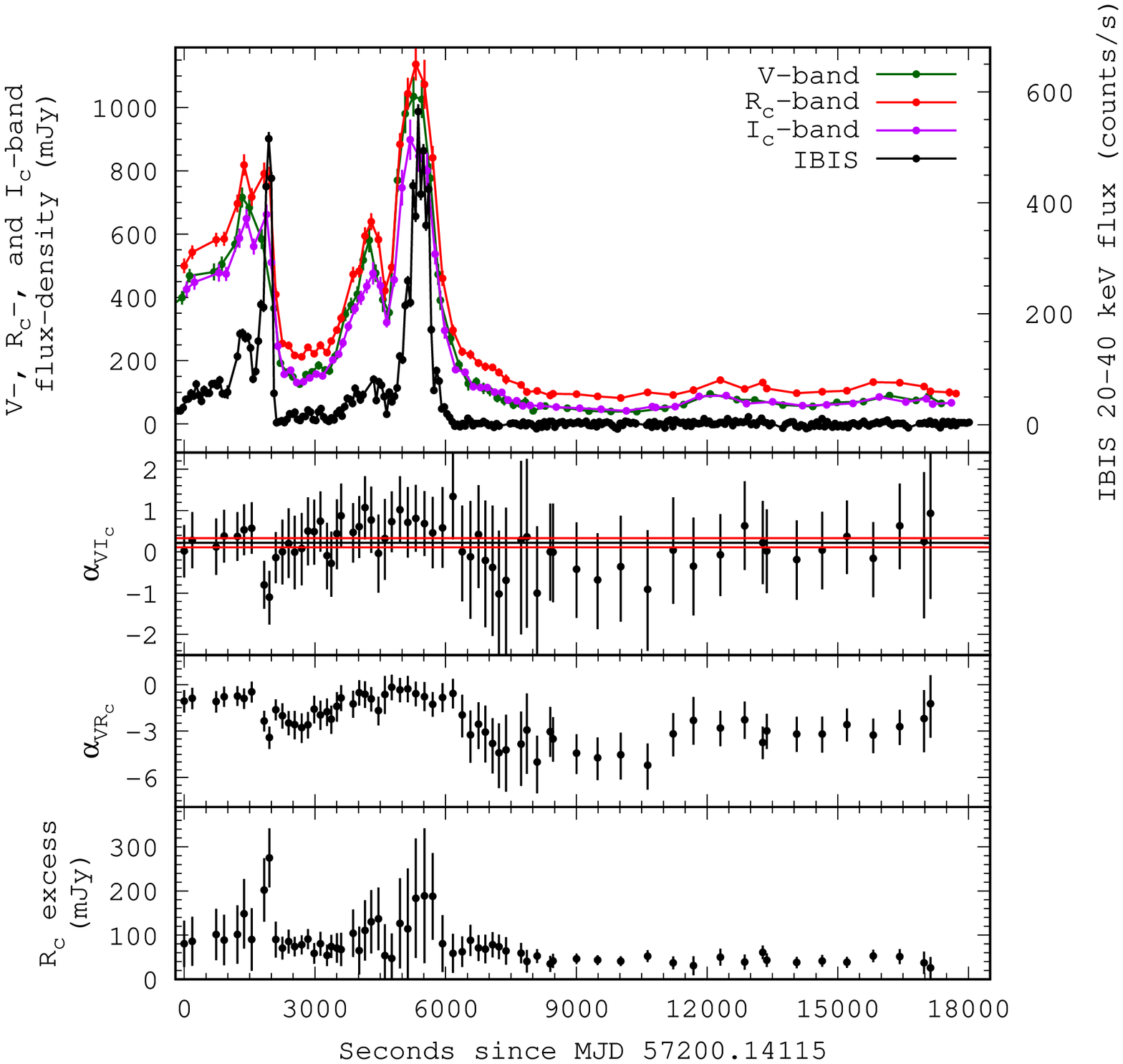} 
   \caption{
   {\it Top panel} shows dereddened, interloper-subtracted \vj-band
   (green), \rc-band (red), and \ic-band (purple) light curves
   obtained by the WCO 0.3m telescope on MJD 57200. The \integral/IBIS
   20--40 keV hard X-ray light curve is shown in black. Note the
   remarkable similarity in the morphology of the optical and the
   hard X-ray light curves.
   {\it Second panel from top}: evolution of the power law slope
   $\alpha_{VI_C}$ connecting \vj-and \ic-bands on the second night.
   There is no significant spectral evolution between the \vj\ and
   \ic\ bands. The best-fit value of $\alpha_{VI_c}$ is 0.22$\pm$0.11
   (assuming $\alpha_{VI_c}$ was constant) is indicated by the solid
   black line and the red dotted lines flanking it.  In a jet
   scenario this would imply that the optically thick to optically thin
   break is at frequencies higher than that of the \vj-band.
   {\it Third panel from top}: evolution of the power law slope
   $\alpha_{VR_C}$ connecting \vj- and \rc-bands. There is significant
   color evolution between the \vj- and \rc-bands, most likely due
   to the presence of a strong \halpha\ emission line contributing
   in the \rc-band.
   {\it Bottom panel}: the \rc-band excess, computed by first
   calculating the interpolated flux at \rc-band based on the
   observed \vj- and \ic-band fluxes (assuming the underlying
   continuum is a power law between \vj\ and \ic), and then subtracting
   this interpolated \rc\ flux from the observed \rc\ flux.
	{\it The WCO and \integral\ light curve data are included with 
	this published article as data behind the figure (DbF) in 
	machine-readable table format.}
   }
 \label{f:n2-lc-slopes-excess} 
\end{figure*} 

\section{Observations and data reduction} \label{s:data} 

\subsection{Wheaton College Observatory data of \src} 
\src\ was observed using Wheaton College Observatory's (WCO; located
in Norton, Massachusetts, USA; latitude = 41$\fdg$965617 N, longitude
= 71$\fdg$184063 W, altitude = 40m) 0.3m {\it Meade LX600-ACF}
Schmidt-Cassegrain reflector telescope during the nights of 2015
June 25-26 local time (MJD 57198), and June 27-28 local time (MJD
57200).  The telescope is equipped with an {\it SBIG STT-8300M}
charge-coupled device (CCD) detector (operated at -20C during our
observations) and standard Johnson-Cousins \uj, \bj, \vj, \rc, and
\ic\ filters.

During the first night (MJD 57198) we repeatedly obtained 30s
exposures in the \vj-band. A 5s delay between successive exposures
was programmed to ensure CCD readout. This was continued throughout
the night, with the exception of a few 1-2 minute-long interruptions
to remove condensation on the telescope's corrector plate. Since
images of \src\ and the reference stars in the field of view are
equally affected by condensation, our results obtained using
differential photometry are not affected. Preliminary photometric
results from this night's observing run were reported in \citet{atel7721}.
Standard optical data reduction processes (bias correction, dark
subtraction, and flat-fielding) were done using {\it astroImageJ}
v3.0.0 \citep{aij_ref}.  Three field stars with AAVSO Unique
Identifiers 000-BCL-467, 000-BCL-468, and 000-BCL-455 were used to
carry out differential photometry of \src, using {\it Aperture
Photometry Tool} v2.4.7 \citep{apt_ref}.  The final \vj-band light
curve, after correcting for interstellar reddening and contribution
from a blended interloper (see \S\ref{s:analysis} below for details),
is shown by green points in Fig.~\ref{f:n1-lc}.

This observing run on MJD 57198 coincided with three \swift\
observations of \src.  The periods of \swift\ overlap are shown by
horizontal lines near the top of Fig.~\ref{f:n1-lc}.  The \swift\
XRT 0.5--10 keV and the UVOT U-band light curves, along with
simultaneous WCO \vj-band light curve, are shown in the bottom three
subpanels of Fig.~\ref{f:n1-lc}.  A summary of the observation logs
is presented in Table~\ref{tab:obslog}.

During the second night (MJD 57200) we cycled through the \vj-,
\rc-, and \ic-band filters throughout the night.  Preliminary
photometric results from this night's observing run were reported
in \citet{atel7737}. As in the first night, we initially started
with 30s exposures in all three bands, but had to increase the
integration times as the source started to fade dramatically after
local midnight. During the latter part of the night, the \vj-band
exposures were five minutes each, and the \rc- and \ic-band exposures
were one minute each.

As indicated in Table~\ref{tab:obslog}, this observing run on MJD
57200 fully coincided with \integral\ revolution number 1557, when
\integral\ was observing \src\ as well.  The \vj, \rc, \ic\ light
curves and simultaneous \integral/IBIS light curves are shown in
Fig.~\ref{f:n2-lc-slopes-excess}.

\begin{deluxetable*}{llcl}
\tabletypesize{\scriptsize}
\tablecaption{Observation Log
\label{tab:obslog}
}
\tablewidth{0pt}
\tablehead{
\colhead{Observation Date} &
\colhead{Observation Duration} & 
\colhead{Instrument and} &
\colhead{Remarks} \\
 \colhead{(UT)} &
 \colhead{(MJD)} &
 \colhead{Bandpass Analyzed} &
 \colhead{}
}
\startdata
2015 Jun 25 & 57198.148 -- 57198.330 & WCO 0.3m; V-band & 440 exposures throughout the 
night, each 30s duration\\
 & & & \\
2015 Jun 25 & 57198.155 -- 57198.160 & \swift; XRT \& UVOT U-band & ObsID:00031403057, visit I\\
2015 Jun 25 & 57198.198 -- 57198.214 & ---''--- & ObsID:00031403057, visit II \\
2015 Jun 25 & 57198.264 -- 57198.275 & ---''--- & ObsID:00031403057, visit III  \\
 & & & \\
2015 Jun 27 & 57200.100 -- 57200.340 & 0.3m; VRI cycled & 64 \vj-band, 66 \rc-band, and 65 \ic-band images \\
 & & & with 30--300s exposures depending on source brightness\\
 & & & \\
2015 Jun 27 & 57200.000 -- 57200.947 & \integral; IBIS 20--40 keV & REV 1557, Observation:1270003/0001\\
\enddata
\end{deluxetable*} 


\subsection{\swift\ and \integral\ observations strictly simultaneous
with WCO} 
While both the \integral\ and \swift\ missions observed \src\
extensively during the outburst, a full analysis of the entire data
set is beyond the scope (and focus) of this current work.  In order
to compare with our optical observations, in this work we only
analyze \integral's IBIS \citep{ibis_ref} 20--40 keV data, \swift's
XRT \citep{xrt_ref} light curve in 0.5--10 keV range, and the U-band
light curve obtained using \swift's UVOT \citep{uvot_ref1,uvot_ref2},
that are strictly simultaneous with our WCO observations. This
includes three \swift\ visits to \src\ on MJD 57198 under ObsID:
00031403057 and the \integral\ observations made during REV 1557
on MJD 57200.

Because the source was very bright, the XRT was operated in
windowed-timing (WT) mode. The data extraction and reduction were
performed using the HEASOFT software \citep[v6.17;][]{heasoft_ref}
developed and maintained by NASA's High Energy Astrophysics Science
Archive Research Center (HEASARC). We followed the extraction steps
outlined in \citet{ReynoldsMiller2013}, and we refer the reader to
that paper for greater details about the XRT and UVOT data reduction.
In summary, the raw XRT data were reprocessed using the {\it
xrtpipeline} tool in order to ensure that the latest instrument
calibrations and responses were used.  Because the XRT data were
collected in WT mode, events were extracted from a rectangular
region containing the source.  Neighboring source-free regions were
used to extract the background properties.  For the UVOT U-band
data, the pipeline-processed level II images were used.  These
images were aspect-corrected using the {\it uvotskycorr} tool, and
then the source flux was extracted from a 5\arcsec\ aperture centered
on the source. The background was extracted from a neighboring
source-free region.  Note that the 5\arcsec\ aperture radius for
the source implies that this also includes the light from the blended
interloper, as in the case of the WCO observations. The count-rate
to flux-density conversion for the UVOT U-band data was carried out
using the appropriate multiplier given in Table 2 of the instrument
calibration document SWIFT-UVOT-CALDB-16-R01\footnote
{\url{http://heasarc.gsfc.nasa.gov/docs/heasarc/caldb/swift}}.

We extracted IBIS/ISGRI \citep{Lebrun+2003} data from revolution
1557 using INTEGRAL Off-line Scientific Analysis 10.2 and following
the standard IBIS/ISGRI cookbook procedure for image, mosaic, and
spectral extractions. Because of the partly very low count rates
of \src\ outside of the flares, we employed the alternative
\textsl{ii\_light} routine for light curve extractions (see
\citet{Grinberg+2011} for a comparison between the different light
curve extraction algorithms) and extracted light curves with a time
resolution of 60\,s, taking into account three bright sources active
in the IBIS field of view (FOV) during the observation (V404 Cyg,
Cyg X--1 and Cyg X--3).



\section{Analysis} \label{s:analysis} 

\subsection{Dereddening and subtraction of flux from a blended interloper} 
\label{ss:interloper}
A star that lies $\sim$1.4\arcsec\ to the north of \src\ \citep{UK1991}
is completely blended in our WCO images.  \citet{Casares+1993}
report the following magnitudes for the blended star: $B = 20.59
\pm 0.05$ mag, $V = 18.90 \pm 0.02$ mag, and $R = 17.52 \pm 0.01$.
\citet{Hynes+2002} suggested an F-type spectrum for the blended
star based on the strong H$\alpha$ absorption feature in its spectra.
\citet{Hynes+2002} also noted that an A$_{\rm V}\sim 4.0\pm0.3$,
i.e. similar to that of \src, is needed for the photometric colors
of the blended star to agree with that of an F-type star.  If the
blended star is an unrelated interloper, there is no reason for it
to have the same extinction as that of \src. However, based on
tables of intrinsic colors of main sequence stars
\citep{intcols_ref1,intcols_ref2}\footnote{Provided online at
\url{http://www.stsci.edu/~inr/intrins.html}}, if the blended star
is an F2 star then it has the same reddening as \src, suggesting a
similar distance. We further noted that an F2V star (of absolute
magnitude $M_V$=2.99; see, e.g., \citealt{PMB2012} and \citealt{PM2013})
extincted by A$_{\rm V}$ = 4.04 (same as that of \src) would appear
to have a \vj-band magnitude of 18.9 if it is at a distance of 2.36
kpc.  Therefore, an F2 star at the same distance as \src\ is consistent
in both color and brightness to those of the blended star. In fact,
this distance estimate to the blended star is so close to that of
the parallax distance to \src\ \citep[2.39 kpc;][]{MJ+2009} that
we speculate in passing whether the blended star is truly unrelated
to the \src\ system, or whether they form a triple system where the
blended star is at least 3,300 au away from \src, orbiting with a
period of at least 50,000 years.

We estimated the flux density of the blended star at other bands
assuming that it is an F2 star, reddened by $A_{V}$ = 4.04.  We note
that this value of $A_{V}$ is consistent with the $A_{V}$ derived
by \citet{Rahoui+2017} based on the equivalent width of the diffuse
interstellar band near 5779 \AA\ observed in a {\it VLT/FORS2}
spectrum obtained between MJD 57200.266--57200.296 (i.e. inside the
time window spanned by our observations on the second night).

The extinction law of \citet{ccm89} was used to compute the dereddened
fluxes.  The ratio of total to selective optical extinction $R_V =
A_V/E(B-V)$ was taken to be 3.1.  The flux from the blended star,
assumed to be constant over our observation timescale, and estimated
using the procedure outlined above, was subtracted from our observed
fluxes in each band.

\subsection{Evolution of the optical colors} \label{ss:colors} 
\begin{figure*}
\centering
\includegraphics[height=0.99\textwidth, angle=-90]{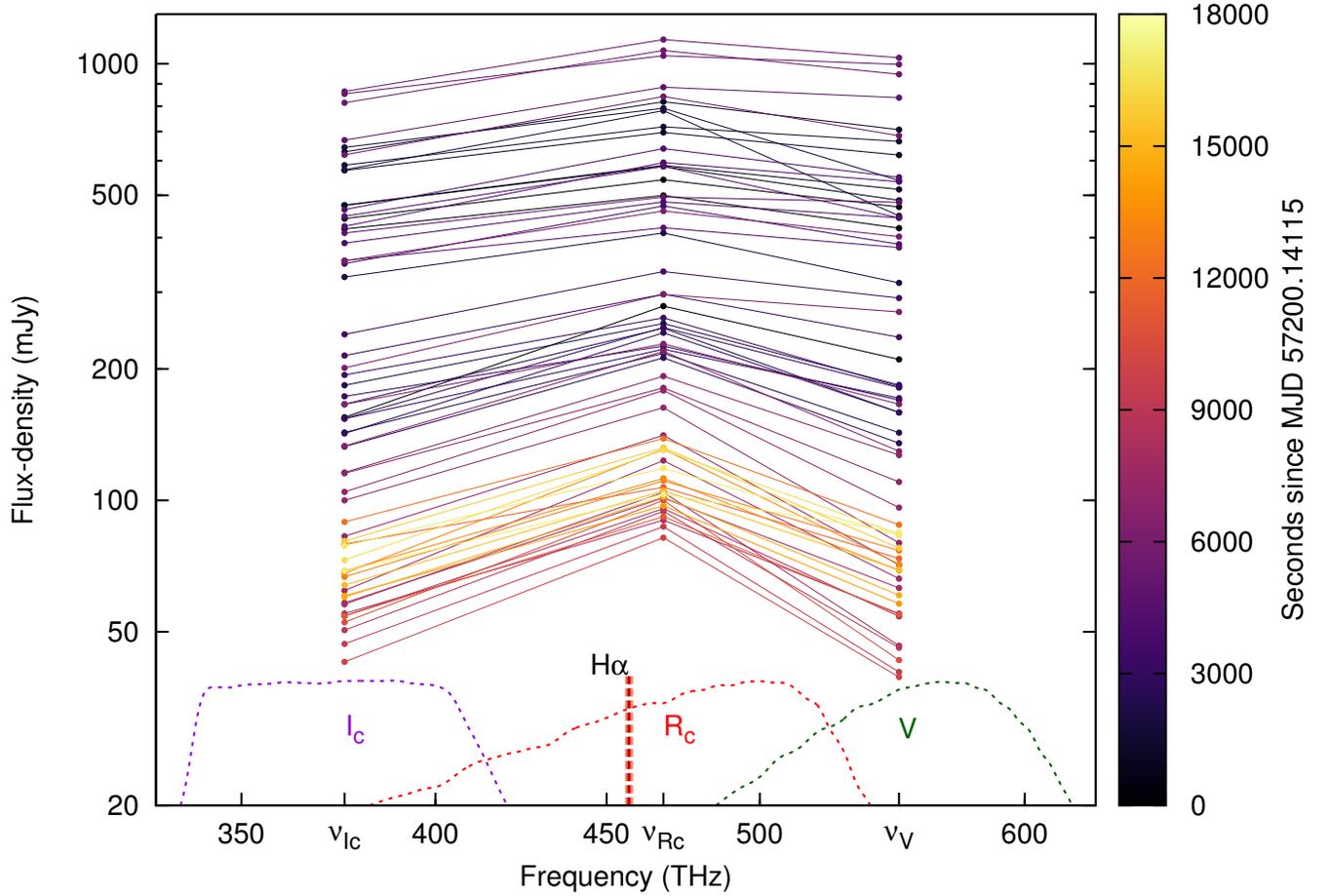} 
   \caption{Evolution of the optical SED during the course of the
   second night's observing run.  This log-log plot clearly shows
   that the SED during the highest overall fluxes is quite flat, whereas
   at low overall brightnesses of \src\ the \vj- and \ic-band fluxes
   drop significantly more than the \rc\ flux, creating a peak in
   the SED near the \rc-band. This excess in the \rc-band, especially
   prominent at low brightness, is most likely due to \halpha\
   emission from a quasi-spherical nebula surrounding the source,
   as suggested by spectroscopic observations reported, e.g., by
   \citet{Gandhi+2016}, \citet{M-D+2016}, and \citet{Rahoui+2017}. The
   location of rest frame \halpha\ emission, as well as our \vj,
   \rc, and \ic\ filter transmission curves, are shown near the
   bottom by dashed lines.
   }
 \label{f:n2-sed} 
\end{figure*}

Pronounced evolution of the \vj--\rc\ color, well correlated with
the source luminosity, was observed during the night of MJD 57200
(i.e. the second night, when we cycled among the \vj\rc\ic\ filters
throughout the night).  This is displayed in the third panel (from
top) of Fig.~\ref{f:n2-lc-slopes-excess}, which shows time variation
of the power law spectral slope between \vj- and \rc-bands. This
is also evident in Fig.~\ref{f:n2-sed}, where evolution of the
dereddened, interloper-subtracted spectral energy distribution (SED)
during the course of the second night's observing run is shown.
Each quasi-simultaneous \vj\rc\ic\ flux triplet is shown by a
connected pair of lines. The color of each triplet indicates the
observation time.  Because we cycled through the filters, the \vj,
\rc, and \ic\ data points were not obtained strictly simultaneously,
and the quasi-simultaneous SEDs were created by choosing observations
closest in time. During the first half of the night, when the source
was bright and flaring, our exposures were 30s in each band and
about 5 s of dead time between images for readout and filter change.
Therefore, spectral changes occurring in timescales less than $\sim$a
minute were not probed during the first half of the night. Similarly,
during the second half of the night when the source became significantly
fainter and less variable, the integration times were increased to
five minutes in \vj\ and one minute each in \rc\ and \ic, implying
that spectral variability occurring in timescales less than $\sim$four
minutes were not probed during the latter half of the night.
Fig.~\ref{f:n2-sed} clearly shows that the power law slope of the
SED during the brightest epochs is quite flat, whereas at low overall
brightnesses the \vj\ and \ic-band fluxes are significantly less
than the \rc-band flux, creating a bump in the SED near the \rc-band.
As discussed in greater detail in the \S\ref{s:conclusion}, the
appearance of  this bump in the \rc-band at lower source luminosities
is most likely due to the presence of a strong \halpha\ emission
line originating in a quasi-spherical nebula surrounding the source
\citep{M-D+2016, Rahoui+2017}.

We also computed the power law spectral slope, $\alpha_{VI_C}$,
between \vj- and \ic-bands. The second panel from the top of
Fig.~\ref{f:n2-lc-slopes-excess} shows the time variation of
$\alpha_{VI_C}$. 
Notice that the value of $\alpha_{VI_C}$ is almost
always close to zero (the best-fit value of $\alpha_{VI_C}$ is
0.22$\pm$0.11). However, both $\alpha_{VR_C}$ and $\alpha_{VI_C}$
tend to approach zero only at highest luminosities.  This spectral
behavior is naturally expected if the \halpha\ line flux from the
nebular emission (which varies on a much slower timescale than the
jet) holds roughly steady as the overall broadband continuum flux
(dominated by the jet) goes down.  The fact that $\alpha_{VI_C}$
largely stays constant with a near-flat slope can be an indication
that (a) these bands trace the optically thick jet continuum, and
(b) the slope of the underlying jet continuum is not strongly
correlated with the luminosity and the slope remains near-flat even
for luminosity changes of $\sim25\times$.

If the underlying broadband emission is indeed due to a jet, then
this would imply that the optically thick to optically thin break
lies at a frequency shorter than that of the \vj-band.  This would
be quite atypical of most X-ray binaries, and is discussed in more
detail in \S\ref{s:conclusion}.  While the average $\alpha_{VI_C}$
slope does not vary much over the course of the night, the errors
on our slope estimation are large enough that we cannot rule out
small ($\sim$0.5--0.8) changes in the slope during individual
observations, especially when the source is faint. For example, the slope
during the lowest flux epoch, about 37 minutes of data centered
around MJD 57200.25, is $\alpha_{\vj\ic} = -0.30^{+0.47}_{-0.41}$.
On the other hand, excluding this low flux interval and keeping the
remaining data yields a slope of $\alpha_{\vj\ic} = 0.25^{+0.11}_{-0.11}$.
Thus, there may be a hint of the slope becoming slightly negative
during epochs of low flux.  The {\it VLT} spectrum obtained during
such an epoch of low flux inside our observation window on MJD 57200
does indeed slope down at the blue end (see \citeauthor{Rahoui+2017}'s
Fig.~10). However, phenomenological modeling of that spectrum also
suggests a break at 3.64$\times$10$^{14}$ Hz \citep{Rahoui+2017},
which is still atypically high for XRBs.  The phenomenon of the
break moving to lower frequencies during epochs of low luminosity,
and the slight anti-correlation between optical polarization and
luminosity found by \citet{Lipunov+2016} and \citet{Shahbaz+2016}
may both be related to the jet's break frequency being directly
related to the energetics of the jet base.  A more energetic jet
base would naturally create the break at higher frequency.

Under the assumption that the broadband continuum emission between
\vj\ and \ic\ is a single power law, the amount of excess emission
in the \rc-band for every \vj\rc\ic\ triplet can be readily evaluated
by subtracting the interpolated \vj--\ic\ power law emission from
the observed \rc-band emission. This \rc-band excess is shown in
the bottom panel of Fig.~\ref{f:n2-lc-slopes-excess}. We note a
slight positive correlation between the \rc-band excess and the
continuum emission.  For example, the \rc\ excess is $\sim$$3$$\times$
larger at the flare peak near MJD 57200.2 compared to that during
the fainter epoch between MJD 57200.25--57200.35.  Note, however,
that the \vj-band flux-density changes by a factor of about 25$\times$
between the abovementioned flare peak and the fainter epoch,
suggesting that the correlation between the \rc\ excess and \vj\
flux-density, while positive, is not linear.

\subsection{Light curve analysis} \label{ss:timing} 
In order to study the light curve morphology and search for time
lags and inherent variability timescales in the data, we computed
the discrete autocorrelation functions (ACFs) and cross-correlation
functions (CCFs) of the light curves using the $z$-transformed
discrete correlation function (ZDCF) method of \citet{zdcf}. The
ZDCF is based on the classic discrete correlation function (DCF)
proposed by \citet{ek1988}; however, unlike the DCF it (a) bins the
data points into equal population bins and (b) employs Fisher's
$z$-transform of the linear correlation coefficient in order to
estimate the confidence level of a measured correlation. Simulations
\citep[see, e.g.,][]{plike} show that the ZDCF provides more reliable
and unbiased error estimation of time lags than the classic DCF
method.  When calculating the ZDCFs we chose to bin a minimum of
11 points per time-lag bin to ensure convergence of the $z$-transform,
omitted zero-lag points in order to avoid spurious peaks, and ran
1000 Monte Carlo simulations for reliable error estimation. The
error on the width of the CCF peak was obtained using a likelihood
method outlined in \citet{plike}, which gives the 68\% fiducial
interval\footnote{As cautioned by \citet{plike}, the concept of a
fiducial interval is different from that of the standard
notion of confidence interval.  Rather, ``[t]he fiducial
interval can be interpreted as the interval where 68\% of the
likelihood-weighted ensemble of all possible CCFs reach their peaks''
\citep{plike}.}.

\begin{deluxetable*}{clcccc}
\tabletypesize{\scriptsize}
\tablecaption{CCF Peak Value, Likelihood at Peak, and Lag estimation
\label{tab:ccf}
}
\tablewidth{0pt}
\tablehead{
\colhead{Observation}            & 
\colhead{Lag Type}               &
\colhead{CCF Peak}               &
\colhead{Likelihood at}          &
\colhead{Lag at CCF}             &
\colhead{68\% Fiducial Interval} \\
 \colhead{}                      &
 \colhead{}                      &
 \colhead{Value}                 &
 \colhead{Peak}                  &
 \colhead{Peak (s)}              &
 \colhead{for Peak (s)}
}
\startdata
MJD~57200                    & \rc-band to \ic-band & 0.99 & 0.83 & -54  & (-162,22)  \\
---''---                     & \vj-band to \rc-band & 0.99 & 0.85 & -58  & (-183,18)  \\
---''---                     & \vj-band to \ic-band & 0.97 & 0.59 & -105 & (-173,29)  \\
---''---                     & 20--40 keV to \vj-band & 0.85 & 0.72 & 54 & (-27,135)  \\
\enddata
\end{deluxetable*} 

\begin{figure}[!b] 
\centering
 \includegraphics[height=0.48\textwidth, angle=-90, clip, trim=-5mm 0mm 0mm -5mm]
 {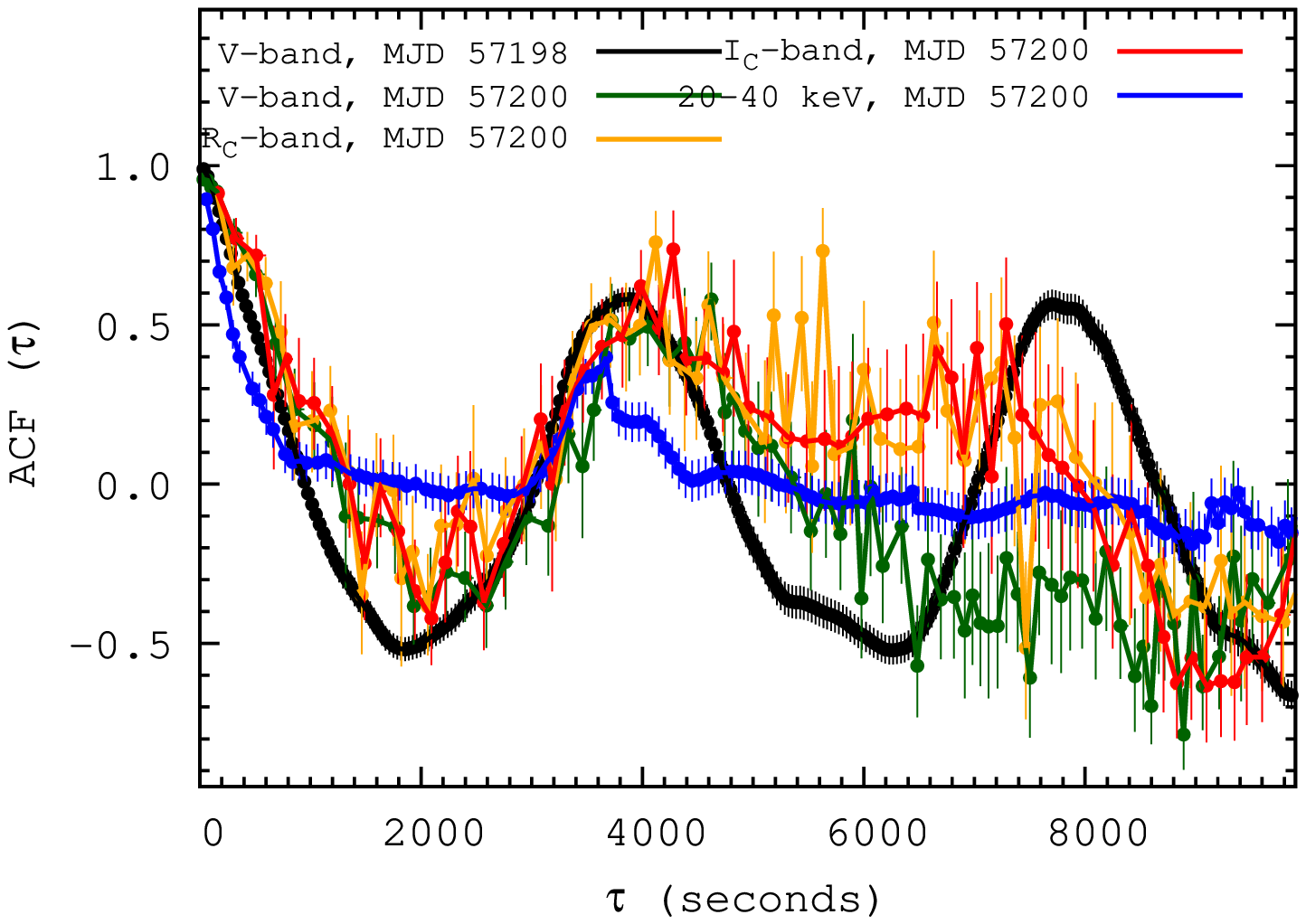} 
   \caption{
   Autocorrelation functions from first night (MJD 57198; \vj-band,
   shown in black), and second night (MJD 57200; \vj-,\rc-, and
   \ic-bands, shown in green, orange, and red, respectively; and
   \integral\ 20--40 keV in blue). Note that apart from the peak
   at zero lag, a second peak at $\sim$4000 s is present in
   all data (optical as well as 20--40 keV X-ray) from both nights.
   A sub-harmonic of the 4000 s peak is evident in the better-sampled
   \vj-band ACF from the first night, and there may be a hint of a
   sub-harmonic peak in the noisier ACFs from the second night.
   Near zero lag, the 20--40 keV X-ray ACF is narrower than the
   optical ACFs, which likely hints at a smaller X-ray emission
   region than the optical.}
 \label{f:acf-all}
\end{figure}

The ACF of first night's \vj-band light curve, shown in black in
Fig.~\ref{f:acf-all}, is remarkably smooth compared to the original
light curve, and apart from the primary peak at zero lag, the ACF
exhibits peaks at $\sim$4000s and at sub-harmonics of the $\sim$4000s
peak.  The \vj-, \rc-, and \ic-band ACFs from the second night (also
shown in Fig.~\ref{f:acf-all}), though noisier than first night's
ACF due to fading of the source as well as sparse sampling, also
show a similar feature at $\sim$4000s.  The morphology of the optical
ACFs may hint that the strong flares are not entirely random, but
might be a manifestation of some limit-cycle type behavior of the
unstable accretion flow \citep[see, e.g.,][]{HKM1991,SM1998, NRM2000}.
In fact, quasi-repetitive, short-term, large-amplitude optical
variability has been reported throughout the bright phase of this
outburst of \src\ by \citet{Kimura+2016}, who attributed this
behavior to disruptions in the inner disk's mass accretion rate.
This claim has recently been supported by contemporaneous optical
spectroscopic observations by \citet{M-D+2016}, who reported strong
outflows of neutral material from the cooler, outer accretion disk.
Analyzing \swift/XRT, {\it Fermi}/GBM, \chandra/Advanced CCD Imaging
Spectrometer (ACIS), {\it NuSTAR},
and {\it INTEGRAL's} IBIS/ISGRI/Joint European X-Ray Monitor (JEM-X) 
data obtained during this
outburst, \citet{Huppenkothen+2017} reported quasi-periodic
oscillations (QPOs) at 18 mHz, 73 mHz, 136 mHz, and 1.03 Hz (i.e.
periods of 55.6 s, 13.7 s, 7.4 s, and 0.97 s respectively) during various
observations. During the previous outburst of \src\ in 1989, optical
QPOs with 210s and 450s periods were reported by \citet{Gotthelf+1991}.
No strong QPOs were found in the {\it Ginga} X-ray data of the 1989
outburst \citep{Oosterbroek+1997}.  There was a period of extreme
flaring near the start of the 1989 outburst of \src\ \citep[see,
e.g., the light curves presented by][]{Wagner+1991}.  However, the data
quality and multiwavelength coverage was poorer in 1989, and hence
it is not clear if the hour-timescale optical variability seen
during the 2015 outburst, seen in our data as well as in the data
presented by \citet{Kimura+2016}, was seen during the 1989 outburst.

The ACF of the \integral/IBIS 20--40 keV light curve that is strictly
simultaneous with our WCO data on MJD 57200 is shown in blue in
Fig.~\ref{f:acf-all}. Similar to the optical ACFs, the 20--40 keV
ACF also shows a peak at $\sim$4000s. The strong sub-harmonic in
the \vj-band data from the first night, peaking near 8000s, is
either absent or highly diminished in the 20--40 keV ACF. At near-zero
lag the X-ray ACF is narrower than the optical, likely suggesting
a smaller hard X-ray emission region than the optical emission
region.

Because the strong optical flux variability without overall color
variability rules out a predominantly disk origin, we suggest that
the disruption in the inner flow leads to a disruption in the
`feeding' of the jet as well. This scenario implies a disk--jet
coupling where the mass outflow rate through the jet is directly
related to the mass inflow rate through the disk.  If other jet
parameters remain largely non-varying, changes in the mass outflow
rate through the jet would cause the \vj\rc\ic\ flux-densities to
change as well, but would not change the spectral shape as long as
all these three optical bands lie in the optically thick part of the
jet synchrotron emission. 

\begin{figure*} 
\centering
 \includegraphics[height=0.65\textwidth, angle=-90, clip, trim=90mm 0mm -20mm 250mm]
 {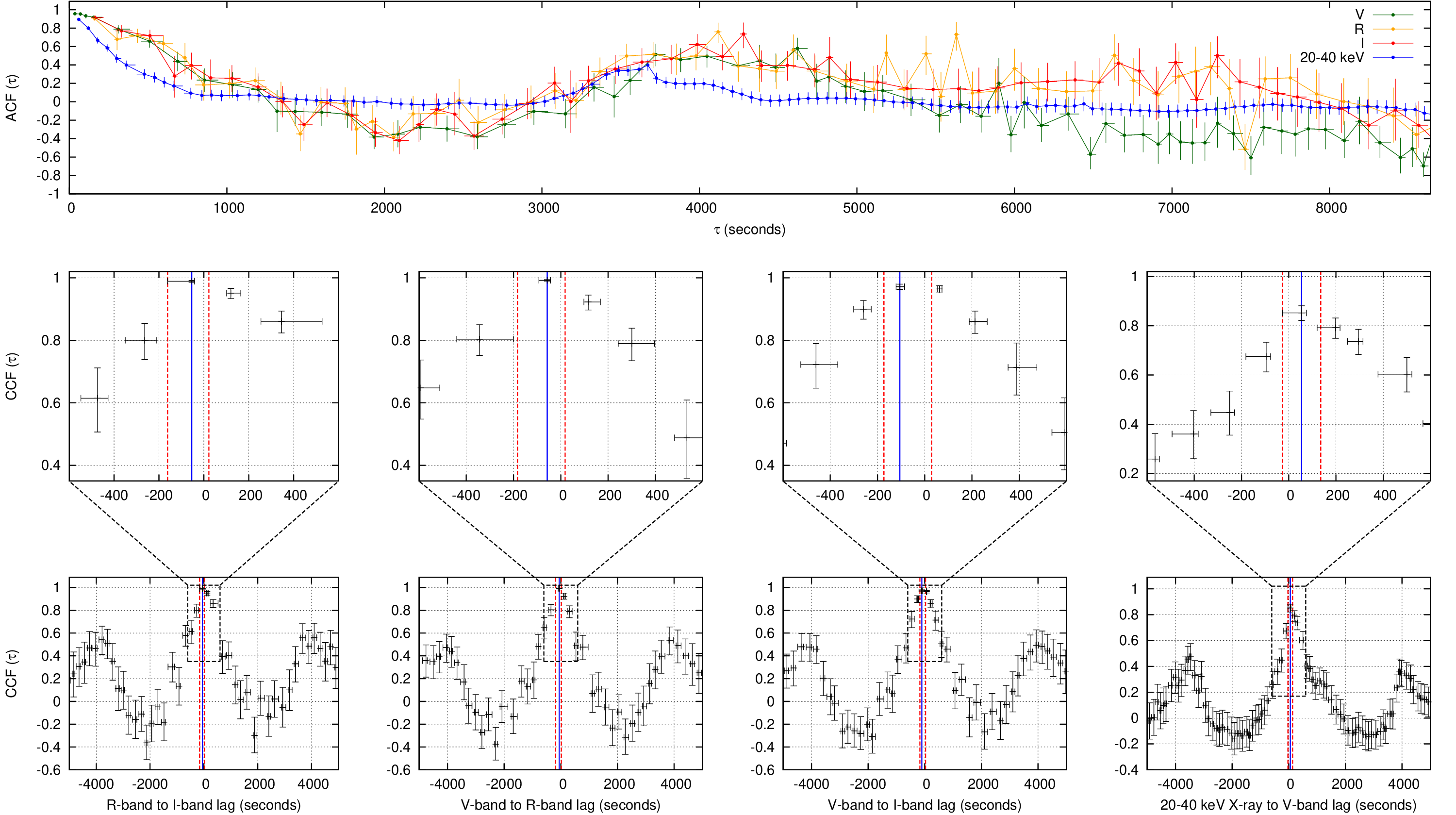} 
 \includegraphics[height=0.65\textwidth, angle=-90, clip, trim=90mm 250mm 0mm 0mm]
 {f5.ps} 
   \caption{
   Cross-correlation functions between the various bands observed
   on the second night, i.e. MJD 57200. The definition of the time
   lag is indicated in the x-axis label of each subpanel.  For each
   plot the vertical blue line indicates the position of the CCF
   peak, and the two red lines flanking the blue line on both sides
   mark the 68\% fiducial interval. See \S\ref{ss:timing} and
   Table~\ref{tab:ccf} for details.  Zoomed insets display CCFs
   near their peaks, showing the behavior of lags between
   $-500\leq\tau\leq500$ s.
   }
 \label{f:n2-ccf}
\end{figure*} 

The CCFs between \rc- and \ic-, \vj- and \rc-, \vj- and \ic, and
20--40 keV and \vj-bands on the second night are shown in
Fig.~\ref{f:n2-ccf}. The location of the CCF peak and the 68\%
fiducial interval for the peak are also indicated.  The error on
the width of the CCF peak, estimated from the 68\% fiducial interval,
suggests that the 20--40 keV X-ray to \vj-band lag is 54$_{-81}^{+81}$s,
the \vj- to \ic-band lag is -105$_{-68}^{+134}$s, the \vj- to
\rc-band lag is -58$_{-125}^{+76}$s, and the \rc- to \ic-band lag
is -54$_{-108}^{+76}$s.  That is, the data are consistent with no
delay between the 20--40 keV, \vj-, \rc-, and \ic-band light curves.
Apart from the main peak, all of the CCFs show a peak at $\tau\sim\pm$4000s.
Properties of the CCFs between the various bands are summarized in
Table~\ref{tab:ccf}.  Strictly simultaneous \swift+WCO data were
obtained on MJD 57198 for only $\sim$nine minutes during visit I (see
Fig.~\ref{f:n1-lc}), for $\sim$21.5 minutes during visit II, and
for $\sim15.5$ minutes during visit III. The relatively short
duration of simultaneity compared to the 35s \vj-band observing
cadence (30s exposures + 5s readout) resulted in noisy CCFs from
which no significant conclusions could be drawn.



\section{Discussion and Conclusions} \label{s:conclusion} 
Based on our observations, and comparing them with those of other
published results of this outburst of \src\ so far, we draw the
following conclusions.
 
\smallskip
(1) The power law slope of the SED between the \vj- and \ic-bands
stays at 0.22$\pm$0.11 during the entire night of our second observing
run, modulo uncertainties in dereddening the fluxes.  For the \vj-
and \ic-band filters used in computing the spectral slope, a 10\%
uncertainty in $A_V$ would lead to a 0.5 uncertainty in the spectral
slope. However, we also note that spectroscopic measurements of
$A_V$ obtained by \citet{Rahoui+2017}, based on the equivalent width
of the diffuse interstellar band near 5779 \AA, observed using
VLT/FOcal Reducer/low dispersion Spectrograph 2 (FORS2) during our
second night's observations, are consistent with previous estimates
of $A_V$$\sim$4 during quiescence. Thus, it seems unlikely that
systematic errors in our spectral slope estimations are significant.
Taken at face value, the average value of the \vj$-$\ic\ slope that
we obtain is marginally consistent with an optically thick accretion
disk, where a slope of 1/3 would be theoretically expected for a
steady-state disk radiating due to viscous dissipation alone.
However, we note that observations of a sample of UV and optical
spectra of short-period black hole X-ray transients in outburst
obtained by \citet{Hynes2005} show that the optical slope is usually
steeper than 1/3 and in the range of $0.5\lesssim\alpha\lesssim1.5$.
Thus, in conjunction with the variability argument presented below,
the \vj--\ic\ slope favors an optically thick jet origin rather
than a disk origin, not to mention the slope is a more natural fit
to what is seen from compact jets in other black hole binaries (see.
e.g. \citet{Stirling+2001} for the case of Cyg X-1).

Because several different emission components of an XRB (e.g. viscous
dissipation, X-ray reprocessing, jet synchrotron) can contribute
significantly in the optical bandpass, it is sometimes possible to
model a flat optical spectrum by adding different amounts of different
components. For example, the SED of XTE J1118+480 obtained by
\citet{Hynes+2000} yielded a flat spectrum with less than 10\%
variation in $f_\nu$ over a factor of 50 in wavelength
($\sim$1000--50000\AA). Nonetheless, the SED could be well described
with a model where the jet break was between optical and NIR, and
a disk component rising into the UV conspired with the jet to create
a flat spectrum \citep{MFF2001,Maitra+2009}. However, such a spectrum
should not be robust during large flux variations, as seen in our
observations of \src. The variability spectrum obtained during a
later (2005) outburst of XTE J1118+480 did indeed suggest a
contribution from optically thin synchrotron \citep{Hynes+2006}.
Thus, in the case of \src, the persistence of the spectrum through
large variability is the key to confirming its origin as an optically
thick jet synchrotron.

The relatively small changes in the amount of linear optical
polarization seen by \citet{Lipunov+2016} and \citet{Shahbaz+2016}
may be an indication that while the jet's break frequency was higher
than that of the \vj-band during most of the time, the break was
not at a much higher frequency. During epochs when the jet base
became less energetically compact, e.g. when the rate at which mass
was fed to the jet decreased and/or when the jet was less magnetically
dominated, the jet break started moving into the optical band, thereby
increasing the amount of polarization.

Additionally, we note that the optical spectral index during the
``slow variations'' (with $\sim$100--1000s timescale) reported by
\citet{Gandhi+2016} during their ULTRACAM observation of this
outburst of \src\ is also similar to what we observe. For the
slow variations, they find a mean spectral index of +0.37$\pm$0.11
between the $u'$ and $g'$ bands and an index of -0.45$\pm$0.09 between
the $g'$ and $r'$ filters (the flux from the \halpha\ line is included
in the slope estimation here; the authors note that correcting for
the \halpha\ will increase the $g'$--$r'$ spectral slope, making it
flatter).

Further indication that the observed optical emission originated in
an optically thick jet comes from comparing simultaneous radio observations.
The 13-18 GHz light curve from the Arcminute Microkelvin Imager-Large Array 
(AMI-LA) published by \citet{Walton+2017} shows that around MJD 57198.3 the
13-18 GHz flux-density from \src\ was $\sim$0.4-0.5 Jy. Our observed \vj-band 
flux at this time was between 3-4 Jy (see Fig.~\ref{f:n1-lc}). This optical 
flux extrapolated to AMI-LA frequencies, assuming a spectral slope of 
$\alpha=0.22$ that we found two nights later, results in an expected AMI-LA
flux of around 0.3-0.4 Jy, which is remarkably close to the AMI-LA flux 
that was actually observed.

\smallskip
(2) No significant changes occurred in the average \vj--\ic\
spectral slope during the $\sim$5 hr of continuous observing on
MJD~57200, even though the flux in individual bands changed by as
much as a factor of $\sim$25. In a jet scenario the easiest way to
reproduce this observation would be the case where the rate at which
matter is fed to the jet's base changes rapidly as well, without
any significant change in the other jet parameters.  The `disrupted
mass flow into the inner regions' scenario based on the detection of
strong outflowing winds from the outer accretion disk
\citep{Kimura+2016,M-D+2016} would naturally account for such a
disrupted feeding of the jet. However, \citet{M-D+2016} found that
from $\sim$ MJD~57200 onwards the P-Cygni profiles were either not
present or extremely weak, suggesting that the dense outflowing winds
stopped, or became optically thin, around that time. The appearance of
many strong optical lines in the spectra after MJD 57200, accompanied
by an increase in the \halpha\ to H\ensuremath{\beta} flux from
$\sim$2.5 to $\sim$6, indicates the formation of a nebular phase
during this time \citep{M-D+2016, Rahoui+2017}.  Our WCO observations
were obtained during the early onset of this nebular phase.

\smallskip
(3) The presence of a strong, slowly varying \halpha\ emission line
was also indirectly inferred in the \rc-band. Strictly simultaneous
spectroscopic  observations of the source reported by \citet{Rahoui+2017},
where a strong \halpha\ emission line was indeed observed, further
strengthen our identification of this spectral feature. In our
broadband filter observations the flux from this line creates a
`bump' in the \rc-band. As shown in Fig.~\ref{f:n2-sed}, this bump
becomes weaker at high luminosities, or, in other words, the line
becomes weaker relative to the continuum.  This anti-correlation
between the line's equivalent width and continuum flux is clearly
visible in the simultaneous $\sim$10 minute-long {\it VLT} spectra
\citep{Rahoui+2017}. While the equivalent width decreases with
increasing source luminosity, the integrated line flux increases
with the luminosity.  Such an anti-correlation has also been noted
during outbursts of other X-ray binaries \citep{Fender+2009}.  In
our data this is shown in the bottom panel of
Fig.~\ref{f:n2-lc-slopes-excess} where the \rc-excess is essentially
a proxy for the integrated line flux. That the increase of the line
flux is slower than the increase in the continuum flux (the line
flux changes only by a factor of $\sim$3 when the continuum flux
changes by a factor of 25 or more) was also noted by \citet{Rahoui+2017}
and attributed to the line being optically thick.

\smallskip
(4) Within the limits of our data, the 20--40 keV X-ray light
curve on MJD 57200 exhibits no lead or lag with respect to the
\vj-band light curve obtained simultaneously. Similarly, we do not
find any lead or lag among the \vj-, \rc-, and \ic-band light curves.
Given that the orbital separation in \src\ is 2.2$\times$10$^7$ km
or $\sim$73 lightseconds, our optical sampling (and hence the DCFs)
is insensitive to delays due to the reprocessing of inner X-rays at the
outer disk or the surface of the donor. Therefore, the absence of
X-ray--optical lag does not rule out an irradiation origin for the
optical. However, the \vj--\ic\ power law slope remained remarkably
constant at 0.22$\pm$0.11 throughout the observing run, insensitive
to flux changes of $\sim25\times$ in the optical (and larger X-ray
flux changes). If the optical were due to irradiation or reprocessing
of the X-rays at the outer disk, then the optical slope would not
have stayed constant. This is because we expect $T_{\rm {irrad}}\propto
\sqrt[4]{L_X}$, where $T_{\rm irrad}$ is the temperature of the
irradiated disk and $L_X$ is the X-ray luminosity; and because changes
in X-ray luminosity were $>$25$\times$, that would have shifted the
irradiation peak by more than a factor of two in wavelength, leading
to observable changes in the optical spectral slope that were not
seen. Thus, the combination of strong variability, lack of time lag,
and lack of optical spectral slope change leads us to suggest that
a dominant fraction of the optical emission originates in a jet.
The stronger variability of the X-rays compared to the optical, and
the narrower ACF of the X-rays, suggests that the X-rays are produced
in a smaller region closer to the central black hole, whereas the
observed optical emission is from the optically thick regions of
the jet located much farther out.

Based on the data presented in this paper, we cannot conclusively
say whetheror not  the X-rays during this epoch were also from a jet. The
X-ray flux from a jet would be directly proportional to the optical
flux in a scenario where both the jet break frequency and the jet's
broadband spectral shape (i.e. spectral slopes above and below the
jet break) does not change. Indeed, as our light curves from the
second night show, the X-ray variations are larger than optical
variations. This may imply that either the X-rays are not coming from a
jet, or perhaps that the X-rays are coming from a jet whose break frequency
and/or its optically thin spectral index is rapidly changing. From what
we know about black hole jets from theory, observations, and modeling,
we expect both the break frequency to change (e.g. due to changing
accretion rate, and/or changing magnetic field strength, and/or
changing the spatial location of particle acceleration regions) and
the optically thin spectral index to change as well (due to changes
in shocks and/or particle acceleration processes).

While observations such as larger X-ray variability than optical
(it would likely require a high-degree of fine tuning a jet model
that produces larger optical variability than X-ray variability),
a very hard X-ray spectrum extending well beyond 100 keV \citep{epline},
and strong reflection features well modeled by a `lamp post' above/blow
the accretion disk \citep{Walton+2017} are nods toward a jet origin
for the X-rays, based on our data alone we cannot strongly constrain
the origin of the X-rays.

\smallskip
If the above conclusions are correct and we were indeed observing
optically thick jet synchrotron at frequencies as high as that of
the \vj-band, the optically thick to optically thin break in the
jet spectrum would be at a frequency even higher  than that of the
\vj-band. This would be extreme in the context of transient XRBs,
where most systems exhibit a break at lower frequencies
\citep{Russell+2013a}.  \citet{Russell+2013a} also found that during
the 1989 outburst of \src\ the break frequency was at (1.8 $\pm$
0.3) $\times$ 10${^{14}}$ Hz, significantly lower than that of the
current outburst.  The SED data used to estimate the jet break
during the 1989 outburst, however, was obtained around MJD 47728--47729.
This was about two months after the 1989 outburst had started, and
by this time the optical/IR flux was more than a factor of 10 lower
than that during the peak of the 1989 outburst. While the coverage
of the 1989 outburst was not as good as that during the 2015 outburst,
\citet{Russell+2013a} noted that a nearly simultaneous radio and
optical-NIR SED obtained on MJD 47676 (near the 1989 outburst peak)
suggested optically thin radio synchrotron and a blue optical-NIR
SED with spectral index $\sim1$. As the fluxes during this time
(i.e. around MJD 47676) were comparable to our 2015 fluxes, better
multiwavelength coverage during this phase of the 1989 would
probably have allowed a better comparison with our 2015 outburst
data.

It is still not clear why the outburst behavior of \src\ is so
different from that of most other XRBs. Dense multiwavelength
monitoring of future outbursts of \src\ and similarly behaved sources
like V4641 Sagittarii (which has a similar, relatively large orbital
period and therefore a large accretion disk as well) and Cygnus
X--3 (which has shown similar high-energy $\gamma$-ray flares),
will be key to unraveling their secrets.


\acknowledgments
We thank the anonymous reviewers for comments and suggestions that
have significantly improved the paper. It is a pleasure to thank Michael
Kahn for a discussion on the pros and cons of various time series analysis 
techniques, and Dr. Greg Schwarz for helping us convert the light curves 
into machine-readable table format, so that the data could be included with the
paper.  We thank Mac Sullivan and John Collins for obtaining the
filter response data using a PerkinElmer Lambda 750 UV/Vis spectrometer
in Wheaton's Laser Spectroscopy Lab.  D.M. and J.S. would like to thank
the Athletics Department at Wheaton for keeping the adjoining
ultraluminous soccer field lights off, which made the WCO observations
possible.

D.M. and J.S. gratefully acknowledge support from NASA Rhode Island
Space Grant Consortium, Wheaton  Mars Student-Faculty Summer Research
Grant and Wheaton Faculty Scholarship Funds.  The research presented
in this paper has made extensive use of data obtained from the
HEASARC data archive, provided by NASA's Goddard Space Flight Center,
NASA's Astrophysics Data System.  We also acknowledge the use of
observations with {\it INTEGRAL}, an ESA project with instruments
and science data center funded by ESA member states (especially the
PI countries: Denmark, France, Germany, Italy, Switzerland, Spain),
and Poland, and with the participation of Russia and the USA.

%
%
\vspace{5mm}
\facilities{
Wheaton College Observatory, 
Swift(XRT),
INTEGRAL(IBIS)
}

\software{
astroImageJ \citep{aij_ref},
Aperture Photometry Tool \citep{apt_ref},
HEASoft \citep{heasoft_ref},
zDCF \citet{zdcf,plike}
}

{\it Note added in proof:} Recent results of a 0.1s delay in the arrival of
the optical light relative to the X-rays have been reported by 
\citet{Gandhi+2017}. Their results indicate that a compact
jet base occasionally (perhaps infrequently) contributes to the fast
variability, and so to the overall emission. Our observations on the
other hand point to a jet component that contributes to the slow
variability, since the WCO data did not probe timescales smaller than
~1 minute. Cross-correlation analysis \citep{Gandhi+2016,Gandhi+2017} shows
that the rapid variations occur simultaneously with the slower
variations. Thus it is possible that the optical jet exhibits two
components simultaneously --- one component that exhibits rapid, red,
optically thin variability, and another component that exhibits
slower, optically thick variations.




\end{document}